\documentclass[12pt,a4paper]{article}
\usepackage[a4paper, margin=1in]{geometry}
\usepackage[normalem]{ulem}
\usepackage{lmodern}
\usepackage[english]{babel}
\usepackage[utf8]{inputenc}
\usepackage[T1]{fontenc}
\usepackage{pdfpages}
\usepackage{microtype}
\usepackage{amsthm}
\usepackage{physics}
\usepackage{xcolor}
\usepackage{hyperref}
\usepackage{bm}
\usepackage{amsmath, amsfonts, amssymb}
\usepackage{authblk}
\usepackage{setspace}
\usepackage{framed}
\usepackage{cancel}
\usepackage{stackengine}
\stackMath
\hypersetup{colorlinks=true,linkcolor=blue,citecolor=blue,urlcolor=blue}
\newcommand{\pbar}{\textbf{\sout{p}}}
\newcommand{\xbar}{\textbf{\sout{x}}}
\newcommand{\alphag}{\boldsymbol{\alpha}}
\newcommand{\betag}{\boldsymbol{\beta}}
\newcommand{\Ze}{\boldsymbol{\zeta}}
\newcommand{\zbar}{\textbf{\sout{z}}}
\newcommand{\Zbbar}{\textbf{\stackinset{c}{}{r}{}{\mbox{--}}{\mathbb{Z}}}}
\newcommand{\p}{\textbf{p}}
\newcommand{\x}{\textbf{x}}

\begin{document}
	\newcommand{\eqrefc}[1]{\textcolor{black}{\eqref{#1}}}
	\newtheorem{theorem}{Proposition}
	\begin{center}
		\textbf{{\Large Casimir operators for the relativistic  quantum\\
				\vspace{0.2cm}
				  phase space symmetry group}}
		\vspace{1cm}\\
		
	\end{center}
\begin{center}
	\textbf{Philippe Manjakasoa Randriantsoa$ ^{1} $, Ravo Tokiniaina Ranaivoson$^{2}$,\\ Raoelina Andriambololona$^{3}$, Roland Raboanary$^{4}$, Wilfrid Chrysante Solofoarisina$^{5}$, Anjary Feno Hasina Rasamimanana$^{6}$.} \vspace{0.5cm}\\
\end{center}

\begin{center}
	\textit{ njakarandriantsoa@gmail.com}$ ^{1} $, 
	\textit{tokiniainaravor13@gmail.com}$ ^{2} $,
	\textit{raoelina.andriambololona@gmail.com}$ ^{3} $ ,
	\textit{r\_raboanary@yahoo.fr}$ ^{4} $, 
	\textit{wilfridc\_solofoarisina@yahoo.fr}$ ^{5} $, 
	\textit{anjaryhasinaetoile@gmail.com}$ ^{6} $ 
\end{center}

	\begin{center}
		$ ^{1,2,3,5,6}$ Institut National des Sciences et Techniques Nucléaires (INSTN-Madagascar)\\
		BP 3907 Antananarivo 101, Madagascar,
		\textit{instn@moov.mg}\vspace{0.3cm}\\
		$ ^{2,3} ${\textit{TWAS Madagascar Chapter, Malagasy Academy,\\BP 4279 Antananarivo 101, Madagascar}}\vspace{0.3cm}\\
		$ ^{1,4, 6} ${\textit{ Faculty of Sciences, iHEPMAD-University of Antananarivo,\\ BP 566 Antananarivo 101, Madagascar}}
	\end{center}

	\begin{abstract}
		Recent developments in the unification of quantum mechanics and relativity have emphasized the necessity of generalizing classical phase space into a relativistic quantum phase space which is a framework that inherently incorporates the uncertainty principle and relativistic covariance. In this context, the present work considers the derivation of linear and quadratic Casimir operators corresponding to representations of the Linear Canonical Transformations (LCT) group associated with a five-dimensional spacetime of signature $(1,4)$. This LCT group, which emerges naturally as the symmetry group of the relativistic quantum phase space, is isomorphic to the symplectic group $Sp(2,8)$. The latter notably contains the de Sitter group $SO(1,4)$ as a subgroup. This geometric setting provides a unified framework for extending the Standard Model of particle physics while incorporating cosmological features. Previous studies have shown that the LCT group admits both fermionic-like and bosonic-like representations. Within this framework, a novel classification of quarks and leptons, including sterile neutrinos, has also been proposed. In this work, we present a systematic derivation of the linear and  quadratic Casimir operators associated with these representations, motivated by their fundamental role in the characterization of symmetry groups in physics. The construction is based on the relations between the LCT group and the pseudo-unitary group $U(1,4)$. Three linears and three quadratics Casimir operators are identified: two corresponding to the fermionic-like representation, two to the bosonic-like representation, and two hybrid operators linking the two representations. The complete eigenvalue spectra and corresponding eigenstates for each operator are subsequently computed and identified .\\

		\textbf{Keywords:} Quantum phase space, Symmetry group, Casimir operator, Linear Canonical Transformations, Beyond standard model physics.
		
	\end{abstract}
	
	\newpage		 
	\section{Introduction}

\quad The concept of Relativistic Quantum Phase Space (QPS) extends classical phase space to relativistic  quantum physics while respecting the uncertainty principle \cite{ravo1,ravo5, ravo2}. Its natural symmetry group is formed by Linear Canonical Transformations (LCTs), which preserve the momenta and coordinates canonical commutation relations. For a spacetime of signature $(N_+, N_-)$, the LCT group is isomorphic to $Sp(2N_+, 2N_-)$ \cite{ravo1}.\\[-0.3cm]

Simultaneously, sterile neutrinos hypothetical right-handed particles with zero Standard Model charges have emerged as promising candidates in the study of physics beyond the Standard Model \cite{ravo2,Naumov,Boyarsky,Drewes,Boser}. Their existence is motivated by neutrino oscillation anomalies, dark matter considerations, and matter-antimatter asymmetry. Recent work has shown that the fermionic representation of LCTs for signature $(1,4)$ naturally suggests sterile neutrino states \cite{ravo2, raoelina}. This signature choice is significant, as it directly relates the algebraic structure to the de Sitter group $SO(1,4)$, offering a potential bridge to the $\Lambda$CDM standard model of cosmology \cite{Mike, Heavens, Martin, Ricardo}.  \\[-0.3cm]

As the LCT group $\mathbb{T} \cong Sp(2,8)$ is non-compact, the systematic study of its unitary representations and their relation to particle classification is not straightforward \cite{harish,rebecca,mackey}. Previous work, however, established a method to address this challenge and identified a framework for obtaining bosonic and fermionic representations, along with their corresponding invariant operators which take the form of quadratic polynomials in the Clifford algebra generators and the coordinates and momenta operators \cite{ravo1,ravo5}. In the present work, our objective is to determine the explicit expression of linear and quadratic Casimir operators associated with these representations. This calculation is complicated by the fact that the Lie algebra corresponding to the fermionic and bosonic representations of the LCTs is not semisimple. Consequently, the standard approach based on the Killing form is not directly applicable. However, this algebra can be shown to be isomorphic to the Lie algebra \( \mathfrak{u}(1,4) \) of the pseudo-unitary group \( U(1,4) \). Although \( \mathfrak{u}(1,4) \) itself is not semi-simple, it admits the following decomposition:

\begin{equation}
	\mathfrak{u}(1,4) \cong \mathfrak{u}(1) \oplus \mathfrak{su}(1,4),
\end{equation}

where \( \mathfrak{su}(1,4) \), the Lie algebra of the special pseudo-unitary group $SU(1,4)$, is simple. This isomorphism between the LCT representations and \( \mathfrak{u}(1,4) \) greatly facilitates the construction of the Casimir operators that is particularly considered in this work.\\

Section 2 establishes the physical and geometric motivation for the five-dimensional $(1,4)$ signature framework, linking it to de Sitter cosmology and sterile neutrinos. Section 3 describes the fermionic (spin) representation of the LCTs and constructs its Casimir operators via the isomorphism with  $\mathfrak{u}(1,4)$. Section 4 presents the derivation of Casimir operators for the bosonic and hybrid representations, completing the set for all three representations, with explicit computation of their eigenvalues and eigenstates on Fock-like states. Finally, Section 5 synthesizes the results, discusses physical implications such as the unification of internal and spacetime symmetries and outlines future research directions.

\newpage

\section{Physical motivation for  the quantum phase space and the five-dimensional $(1,4)$-signature  framework}

\subsection{Relativistic quantum phase space and saturation of uncertainty relations}

The extension of the notion of phase space from classical to quantum physics is non-trivial due to the uncertainty relations that arise from the momenta and coordinates Canonical Commutation Relations (CCRs). The concept of relativistic Quantum Phase Space (QPS) provides a solution by defining basic phase space states that respect these quantum constraints \cite{ravo1, ravo5, ravo2}. 

The Quantum Phase-Space (QPS) formalism introduces fundamental states that achieve equality in the  uncertainty relations. The coordinates wavefunctions  associated to these states, denoted $ z = \lvert\{ \lvert z_{\mu}\rangle\} \rangle$ are gaussian-like functions \cite{ravo2} :
\begin{equation}
	\langle \mathbf{x}^{\mu} | \{(z_{\mu})\} \rangle = \langle x \lvert \langle z \rangle \rangle = \frac{e^{\frac{-B_{\mu\nu}}{\hbar^2}({x}^{\mu}-\langle {x}^{\mu}\rangle)({x}^{\nu}-\langle {x}^{\nu}\rangle) - \frac{i}{\hbar}\langle p_{\mu}\rangle {x}^{\mu} + iK}}{\left| (2\pi)^N \left| \det[\mathcal{X}^{\mu}_{\nu}] \right| \right|^{1/4}},
\end{equation}
where $\langle p_{\mu}\rangle = \langle (z) | \mathbf{p}_{\mu} | (z) \rangle$ and $\langle x_{\mu}\rangle = \langle (z) | \mathbf{x}_{\mu} | (z) \rangle$ are expectation values of momenta and coordinates operators, and $B_{\mu\nu}$ are parameters related to the momenta-coordinates variance-covariance matrix.

These states $|(z)\rangle$ are eigenstates of the operator \cite{ravo1, ravo5, ravo2}:
\begin{equation}
	\mathbf{z}_{\mu} = \mathbf{p}_{\mu} + \frac{2i}{\hbar} B_{\mu\nu} \mathbf{x}^{\nu},
	\label{eq:defz}
\end{equation}

The corresponding eigenvalue equation is :

\begin{equation}
	\mathbf{z}_{\mu} |(z)\rangle = \left[ \langle p_{\mu}\rangle + \frac{2i}{\hbar} B_{\mu\nu} \langle\mathbf{x}^{\nu}\rangle \right] |(z)\rangle = \langle z \rangle  \ket{\langle z \rangle }
	\label{eq:z}.
\end{equation}

The parameters $B_{\mu\nu}$ are explicitly given by \cite{ravo5}:
\begin{equation}
	B_{\mu\nu} = \frac{\hbar^2}{4} (\eta_{\mu\rho} + \frac{2i}{\hbar} \varrho_{\mu\rho}) \tilde{\mathcal{X}}^{\rho}_{\nu},
\end{equation}

where $\tilde{\mathcal{X}}^{\rho\nu}$ are parameters directly related to the components $\mathcal{X}_{\mu\nu}$, $\mathcal{P}_{\mu\nu}$, and $\varrho_{\mu\nu}$ of the momenta-coordinates variance-covariance matrix $ \begin{pmatrix}
	\mathcal{P} & \varrho \\ \varrho^{\tau} & \mathcal{X}
\end{pmatrix}$  associated with the state  $|\langle z\rangle\rangle$. 

The quantum phase space can then be defined as the set $\{\langle z_{\mu} \rangle \}$ of all possible values of the expectation values of the operators $\mathbf{z}_{\mu}$, or equivalently as the set $\{(\langle p_{\mu}\rangle, \langle x_{\mu}\rangle)\}$ of possible expectation values of momenta and coordinates operators pairs for a given variance-covariances structure \cite{ ravo5, ravo2}. This explicit dependence on quantum uncertainties makes the QPS fundamentally compatible with the uncertainty principle.

\subsection{The signature $(1,4)$ and sterile neutrinos}

The choice of signature $(1,4)$ is particularly significant for both cosmology and  particle physics. In fact, this signature corresponds  to the symmetry group $SO(1,4)$ of the de Sitter relativity \cite{ravo1, ravo2,Mike,sitter1,sitter2,sitter3,sitter4}. De Sitter space describes a vacuum with positive cosmological constant $\Lambda$, matching the current $\Lambda$CDM model of cosmology \cite{Mike, Heavens, Martin, Ricardo}. The de Sitter radius $R_{dS} = \sqrt{3/\Lambda}$ introduces a fundamental length scale that may naturally explain the smallness of neutrino masses. In our framework, the de Sitter group $SO(1,4)$ emerges as a subgroup of the LCT group $ \mathbb{T} \cong Sp(2,8)$, providing a geometric basis for unifying spacetime transformations with internal quantum properties \cite{ravo1}.\\

Sterile neutrinos are right-handed singlet states with all Standard Model charges equal to zero: weak isospin $I_3 = 0$, weak hypercharge $Y_W = 0$, electric charge $Q = 0$, and strong color charges zero. The introduction of sterile neutrinos addresses multiple challenges in contemporary physics \cite{ravo2,Naumov,Boyarsky,Drewes,Boser,raoelina}:

\begin{itemize}
	\item Neutrino oscillation anomalies: Experiments have reported anomalies suggesting additional neutrino states beyond the three active flavors.
	
	\item Neutrino masses: Within the Standard Model, neutrinos are massless, yet oscillation experiments demonstrate they have non-zero masses. Sterile neutrinos can facilitate seesaw mechanisms or other mass-generation schemes .
	
	\item Matter-antimatter asymmetry: Additional CP-violation sources from sterile neutrino sectors could explain the baryon asymmetry of the universe.
	
	\item Dark matter: Certain sterile neutrino mass ranges make them viable warm dark matter candidates.
	
	\item Cosmological concordance: The de Sitter framework with signature $(1,4)$ naturally incorporates a positive cosmological constant, consistent with observational cosmology.
\end{itemize}

 In the  formulation given in \cite{ravo1, ravo5, ravo2}, the sterile neutrino states emerge naturally from the spin representation of LCTs for signature $(1,4)$. This geometric origin distinguishes sterile neutrinos from ad hoc extensions of the Standard Model, providing a principled explanation for their existence and quantum numbers.\\

\section{Casimir operators of  the fermionic representation }

\subsection{Special pseudo-orthogonal representation of LCTs}

Linear Canonical Transformations (LCTs) form the natural symmetry group of the Relativistic Quantum Phase Space. They are defined as linear transformations mixing momentum and coordinate operators while preserving the canonical commutation relations \cite{ravo1}:
\begin{equation}
	\begin{cases}
		\mathbf{p}'_\mu = \mathbb{A}^\nu_\mu \mathbf{p}_\nu + \mathbb{B}^\nu_\mu \mathbf{x}_\nu \\
		\mathbf{x}'_\mu = \mathbb{C}^\nu_\mu \mathbf{p}_\nu + \mathbb{D}^\nu_\mu \mathbf{x}_\nu
	\end{cases}
\end{equation}
subject to : 

\begin{equation}
	\begin{cases}
		\Big{[}\p^{'}_{\mu},\x^{'}_{\nu}\Big{]} = \Big{[}\p_{\mu},\x_{\nu}\Big{]} = i \hbar\eta_{\mu\nu}\\[0.12cm]
		\Big{[}\p^{'}_{\mu},\p^{'}_{\nu}\Big{]} = \Big{[}\p_{\mu},\p_{\nu}\Big{]} = 0 \\[0.12cm]
		[\x^{'}_{\mu},\x^{'}_{\nu}] = [\x_{\mu},\x_{\nu}] = 0
	\end{cases}
	\label{eq:lcts}
\end{equation}
This establishes the isomorphism $\mathbb{T} \cong Sp(2, 8)$ between the LCT group and the symplectic group for  the spacetime signature $(1,4)$ \cite{ravo1, ravo5, ravo2}.\\

The variance-covariance matrix can be factorised in the following form:

\begin{equation}
	\begin{pmatrix}
		\mathcal{P} & \varrho \\  \varrho^{\tau}  & \mathcal{X} \end{pmatrix} = 	\begin{pmatrix}
			b & 0 \\  2 acb & a
		\end{pmatrix}^{\tau} \begin{pmatrix}
		\eta & 0 \\  0 & \eta 	\end{pmatrix} \begin{pmatrix}
			b & 0 \\  2 acb & a
	\end{pmatrix}
\end{equation}

Taking advantage of this factorisation, we can introduce the  reduced momenta and coordinates operators defined by the following relations:

\begin{equation}
	\begin{cases}
		\pbar_\mu = \sqrt{2}\, a^{\nu}_{\mu}\,(p_\nu - \langle p_\nu \rangle)
		- \sqrt{2}\, c^{\nu}_{\mu}\,(x_\nu - \langle x_\nu \rangle), \\[6pt]
		\xbar_\mu = \sqrt{2}\, b^{\nu}_{\mu}\,(x_\nu - \langle x_\nu \rangle)
	\end{cases}	
		\label{eq:reduce}
\end{equation}

The transformation laws of these operators are \cite{ravo5}: 

\begin{equation}
	\begin{cases}
		\pbar^{'}_\mu = \Pi^{\nu}_{\mu}\, \pbar_\nu + \Theta^{\nu}_{\mu}\, \xbar_\nu, \\[6pt]
		\xbar^{'}_\mu = -\Theta^{\nu}_{\mu}\, \pbar_\nu + \Pi^{\nu}_{\mu}\, \xbar_\nu
	\end{cases}
	\quad \Longleftrightarrow \quad
	(\pbar^{'} \;\; \xbar^{'}) = (\pbar \;\; \xbar)
	\begin{pmatrix}
		\Pi & -\Theta \\
		\Theta & \Pi
	\end{pmatrix}
		\label{eq:reduce1}
\end{equation}

With $	\begin{pmatrix}
	\Pi & -\Theta \\
	\Theta & \Pi
\end{pmatrix}$ a matrix that is both symplectic and pseudo-orthogonal, i.e, it is an element of the group $\mathbb{G}$ which satisfies the following relation: 
\begin{equation}
	\mathbb{G} \cong Sp(2,8) \cap O(2,8)\cong Sp(2,8) \cap SO_0(2,8) \cong U(1,4)
\end{equation}

The transformation laws for the reduced  operators thus defines a special pseudo-orthogonal representation of the LCTs which can be described explicitily by the specification of a surjective homomorphism $\mathfrak{f}$ from the LCT group $\mathbb{T}$ to $\mathbb{G}$.

\begin{equation}
	\mathfrak{f} : \mathbb{T} \longrightarrow \mathbb{G} \cong \mathbb{T}/ \mathbb{H}
\end{equation}
\begin{equation}
	\quad \begin{pmatrix}
		 \mathbb{A} & \mathbb{C} \\
		 \mathbb{B} & \mathbb{D}
	\end{pmatrix}
	\longmapsto
	\begin{pmatrix}
		\Pi & -\Theta \\
		\Theta & \Pi
	\end{pmatrix} \qquad
\end{equation}

In which $\mathbb{H}$ is the normal subgroup of $\mathbb{T}$ formed by the LCTs that leave the reduced operators invariants: $\mathbb{H} = Ker(\mathfrak{f})$ \cite{ravo1}.

\subsection{Spin representation of LCTs and its Casimir operators}

The special pseudo-orthogonal representation via the group $\mathbb{G}$ allows us to define the spin representation \cite{Porteous95,Lounesto01,BennTucker87} of LCTs with the introduction of a covering map \textit{u}.

   \begin{equation}
	\begin{cases}
		\textit{u} : \mathbb{S} \longrightarrow \mathbb{G} \\[0.3cm]
		\hspace{0.5cm} \mathcal{S} \longrightarrow  \begin{pmatrix}  \Pi & - \Theta \\ \Theta & \Pi \end{pmatrix}
	\end{cases}
	\iff
	\begin{cases}
		\begin{pmatrix}  {\pbar}^{'} & {\xbar}^{'} \end{pmatrix}  = \begin{pmatrix}  \pbar & \xbar \end{pmatrix} \begin{pmatrix}  \Pi & - \Theta \\ \Theta & \Pi \end{pmatrix} \\[0.3cm]
		\hspace{1.1cm}	\Zbbar^{'} = \mathcal{S} \Zbbar \mathcal{S}^{-1} 
	\end{cases}
\end{equation}

In which:
 \begin{itemize}
 	\item $\mathbb{S}$ is the topological double cover of $\mathbb{G}$. It is a subgroup of the spin group $Spin(2,8)$.
 	\item  $\mathcal{S}$ and $- \mathcal{S}$  are the elements of $\mathbb{S}$ which corresponds to the LCTs associated to the matrix $\begin{pmatrix}  \Pi & - \Theta \\ \Theta & \Pi \end{pmatrix} $ : $Ker(\textit{u}) \cong (1, -1)$.
 	\item $\Zbbar$ is the hybrid operator defined by the following relation: 
 	\begin{equation}
 	\Zbbar = 	\cfrac{1}{\sqrt{2}}	\big{(}\alphag^{\mu}\pbar_{\mu} +  \betag^{\mu}\xbar_{\mu}\big{)}
 	\label{eq:hybrid}
 	\end{equation} 
 	with $\alphag^{\mu}$ and $\betag^{\mu}$ the generators of the Clifford algebra $Cl(2,8)$ :
 \end{itemize}
    
    \begin{equation}
  	\begin{cases}
  		\alphag^{\mu}\alphag^{\nu} + \alphag^{\nu}\alphag^{\mu} = 2\eta^{\mu\nu}\\
  		\betag^{\mu}\betag^{\nu} + \betag^{\nu}\betag^{\mu} = 2\eta^{\mu\nu}\\
  		\alphag^{\mu}\betag^{\nu} + \betag^{\nu}\alphag^{\mu} = 0
  	\end{cases}
  	\label{eq:abeta}
  \end{equation}\\

We may introduce the following operators  : 

\begin{equation}
	\left\{
	\begin{aligned}
		\zeta^{\mu} &= \frac{1}{2}\left(\alphag^{\mu} + i\,\betag^{\mu}\right), \\[4pt]
		\zeta^{\mu *} &= \frac{1}{2}\left(\alphag^{\mu} - i\,\betag^{\mu}\right)
		= \zeta^{\dagger}_{\mu}, \\[4pt]
		\zeta^{\mu \dagger} &= \frac{1}{2}\left(\alphag^{\mu \dagger} - i\,\betag^{\mu \dagger}\right)
		= \zeta^{*}_{\mu}
	\end{aligned}
	\right.	
	\label{eq:zeta}
\end{equation}

It can be deduced from (\ref{eq:abeta}) and (\ref{eq:zeta}) that these operators satisfy the following anticommutation relations: 

  \begin{equation}
	\begin{cases}
	\Ze^{\mu}\Ze^{\nu} + \Ze^{\nu}\Ze^{\mu} = 0\\
	{\Ze^{\mu}}^{*}{\Ze^{\nu}}^{*} + {\Ze^{\nu}}^{*}{\Ze^{\mu}}^{*} = 0\\
	{\Ze^{\mu}}^{*}\Ze^{\nu} + \Ze^{\nu}{\Ze^{\mu}}^{*} = \eta^{\mu\nu}
 \end{cases}
	\iff
	\begin{cases}
		\Ze^{\mu}\Ze^{\nu} + \Ze^{\nu}\Ze^{\mu} = 0\\
		{\Ze^{\mu}}^{\dagger}{\Ze^{\nu}}^{\dagger} + {\Ze^{\nu}}^{\dagger}{\Ze^{\mu}}^{\dagger} = 0\\
		{\Ze^{\mu}}^{\dagger} \Ze^{\nu}+ \Ze^{\nu}{\Ze^{\mu}}^{\dagger} = \delta^{\mu\nu}
	\end{cases}
	\label{eq:anticom}
\end{equation}

Then, it can be shown that a basis of the Lie algebra $\mathfrak{s}$ of the group $\mathbb{S}$ is the family $\{\bm{\Xi}^{\mu \nu} \} $  defined by the following relation:

\begin{equation}
	\bm{\Xi}^{\mu \nu} = \cfrac{1}{2}( {\Ze^{\mu}}^{*}{\Ze^{\nu}} - \Ze^{\nu}\Ze^{\mu *} ) = {\Ze^{\mu}}^{*}{\Ze^{\nu}} - \cfrac{1}{2} \eta^{\mu \nu} 
	\label{eq:xi}
\end{equation}

The associated commutation relation is:

\begin{equation}
	[\bm{\Xi}^{\mu \nu}, \bm{\Xi}^{\rho \sigma}  ]= \eta^{\nu \rho}\bm{\Xi}^{\mu \sigma} - \eta^{\mu \sigma}\bm{\Xi}^{\rho \nu} 
\end{equation}

This commutation relations permit to verify that $\mathfrak{s}$ is isomorphic to the algebra $\mathfrak{u}(1,4)$, this can also be seen as a consequence of the group isomorphism $\mathbb{G} \cong U(1,4)$. This fact allow us to deduce that the linear and quadratic Casimir operators denoted respectively $\mathcal{C}^{(1)}_{F}$ and $\mathcal{C}^{(2)}_{F}$ are given by the following relation.

\begin{equation}	
    \begin{cases}
		\bm{\mathcal{C}}^{(1)}_{F} = \eta_{\mu \nu} \bm{\Xi}^{\mu \nu} \\
		\bm{\mathcal{C}}^{(2)}_{F} =  \eta_{\mu \rho} \eta_{\nu \sigma} \bm{\Xi}^{\mu \nu} \bm{\Xi}^{\rho \sigma}
	\end{cases}
	\label{eq:fcasimir}
\end{equation}

Like in \cite{ravo2, ravo5} we consider the  operators :  $\bm{\Sigma}^{\mu \nu} = {\Ze^{\mu}}^{\dagger} \Ze^{\nu}$ and $ \bm{\Sigma} = \delta_{\mu\nu} \bm{\Sigma}^{\mu \nu}$. The operators ${\Ze^{\mu}}^{\dagger}$ and $\Ze^{\mu}$ act as ladder operators for  $\bm{\Sigma}^{\mu \mu}$ and $\bm{\Sigma}$. Their eigenvalues equations are: 

\begin{equation}
  \begin{cases}
  	\bm{\Sigma}^{\mu \mu} \lvert f \rangle =  \bm{\Sigma}^{\mu \mu} \lvert f^{0}, f^{1},f^{2},f^{3},f^{4} \rangle = f^{\mu} \lvert f^{0}, f^{1},f^{2},f^{3},f^{4} \rangle =  f^{\mu} \lvert f \rangle \\
  	\bm{\Sigma} \lvert f \rangle = (f^{0}+ f^{1}+f^{2}+f^{3}+f^{4})\lvert f \rangle  = \lvert f \lvert  \ket{f}
  \end{cases}
  \label{eq:eign}
\end{equation}  

We have between $\bm{\Xi^{\mu \nu}}$ and $\bm{\Sigma^{\mu \nu}}$ the following relation: 

\begin{equation}
	\begin{cases}
		\bm{\Sigma}^{00} = \bm{\Xi}^{00} + \cfrac{1}{2} \\
		\bm{\Sigma}^{0j} = \bm{\Xi}^{0j}\\
		\bm{\Sigma}^{j0} = \bm{\Xi}^{j0} \\
		\bm{\Sigma}^{jj} = \bm{\Xi}^{jj}+ \cfrac{1}{2}\\
		\bm{\Sigma}^{jk} = - \bm{\Xi}^{jk} \quad (j\neq k)
	\end{cases}
	\iff	
	\begin{cases}
		\bm{\Xi}^{00} =  \bm{\Sigma}^{00} - \cfrac{1}{2} \\
	    \bm{\Xi}^{0j}	 =   \bm{\Sigma}^{0j}\\
		 \bm{\Xi}^{j0} = \bm{\Sigma}^{j0}\\
	     \bm{\Xi}^{jj} = 	\bm{\Sigma}^{jj}- \cfrac{1}{2}\\
		 \bm{\Xi}^{jk} = - \bm{\Sigma}^{jk}  \quad (j\neq k)
	\end{cases}
	\label{eq:xisig}
\end{equation}

From  relations (\ref{eq:fcasimir}), (\ref{eq:eign}) and (\ref{eq:xisig}) we can deduce that the eignevalues equations of the Casimir operators are: 

\begin{equation}
	\begin{cases}
		\bm{\mathcal{C}}^{(1)}_{F} \lvert f \rangle =  \Big{(}\bm{\Sigma} - \cfrac{5}{2} \Big{)} \lvert f\rangle = \Big{(}\lvert f \lvert - \cfrac{5}{2} \Big{)} \lvert f\rangle\\
		\bm{\mathcal{C}}^{(2)}_{F} \lvert f \rangle =  \cfrac{5}{4}\lvert f \rangle 
	\end{cases}
	\label{eq:eignf}
\end{equation}  

The operators $\bm{\mathcal{C}}^{(1)}_{F} $ and $\bm{\mathcal{C}}^{(2)}_{F}$  have the same eigenstates $\lvert f \rangle $ as the operators  $\bm{\Sigma}^{\mu \mu}$ and  $\bm{\Sigma}$, and their eigenvalues are respectively $( \lvert f \lvert - \frac{5}{2})$  and  $\frac{5}{4}$.

\section{ Casimir operators of  the bosonic and hybrid representations }

\subsection{Algebra and Casimir operators associated to the bosonic-like representation}

From the relation (\ref{eq:defz}), (\ref{eq:z}) and (\ref{eq:reduce}) we may introduced the following operators:

\begin{equation}
	\begin{cases}
		\bm{\zbar}_\mu =  a_\mu^{\nu}\,(\bm{z}_\nu - \langle z_\nu \rangle)  =  \dfrac{1}{\sqrt{2}}\,(\pbar_\mu + i\,\xbar_\mu)
		\\[8pt]
		
		\bm{\zbar}_\mu^{\star} = a_\mu^{\nu}\,(\bm{z}_\nu^{\dagger} - \langle z_\nu \rangle^{*}) = \dfrac{1}{\sqrt{2}}\,(\pbar_\mu - i\,\xbar_\mu)
		= \zbar^{\mu\dagger} \\[8pt]
		
		\zbar_\mu^{\dagger} = a_\mu^{\nu *}\,(\bm{z}_\nu^{\dagger} - \langle z_\nu \rangle^{*}) = \dfrac{1}{\sqrt{2}}\,(\pbar_\mu^{\dagger} - i\,\xbar_\mu^{\dagger})
		= \bm{z}^{\mu^{\star}}
	\end{cases}
	\label{eq:zmu}
\end{equation}

The  commutation relations for these operators are: 

 \begin{equation}
	\begin{cases}
		\zbar_{\mu}\zbar_{\nu} - \zbar_{\nu}\zbar_{\mu} = 0\\
		{\zbar_{\mu}}^{*}{\zbar_{\nu}}^{*} - {\zbar_{\nu}}^{*}{\zbar_{\mu}}^{*} = 0\\
		{\zbar_{\mu}}^{*}\zbar_{\nu} - \zbar_{\nu}{\zbar_{\mu}}^{*} = \eta_{\mu\nu}
	\end{cases}
	\iff
	\begin{cases}
	\zbar_{\mu}\zbar_{\nu} - \zbar_{\nu}\zbar_{\mu} = 0\\
	{\zbar_{\mu}}^{\dagger}{\zbar_{\nu}}^{\dagger} - {\zbar_{\nu}}^{\dagger}{\zbar_{\mu}}^{\dagger} = 0\\
	{\zbar_{\mu}}^{\dagger}\zbar_{\nu} - \zbar_{\nu}{\zbar_{\mu}}^{\dagger} = \delta_{\mu\nu}
\end{cases}
	\label{eq:com}
\end{equation}

The transformation laws of the operators $\zbar_{\mu}$ can be deduced from the relations (\ref{eq:reduce}) and (\ref{eq:zmu}), one obtains: 

\begin{equation}
\zbar^{\mu^{'}} = (\bm{\Pi}^{\nu_{\mu}} - i \bm{\Theta}^{\nu}_{\mu} ) \bm{\zbar_}{\mu} \iff \bm{\zbar^{'}} = \bm{\zbar}(\Pi - i \Theta)
\end{equation} 

In which the  $5\times 5$ square matrix  $(\Pi - i \Theta)$ belongs to the group $U(1,4) \cong \mathbb{G} \cong Sp(2,8) \cap SO_0(2,8)$ i.e. it verifies the relation  $(\Pi - i \Theta)^{\dagger} \eta (\Pi - i \Theta) = \eta $. \\

The representation laws of operator $\zbar^{\mu} $ can be put in the form:  $\bm{\zbar_{\mu}^{\prime}}
=
\bm{U}^{\dagger}\,\bm{\zbar_{\mu}}\,\bm{U}$ with $\bm{U}$ also verifying $\bm{U}^{\dagger}\,\eta\,\bm{U} = \eta$. $\bm{U}$ itself can be written in the form $\bm{U} = e^{\vartheta^{\mu \nu} \bm{\Upsilon_}{\mu\nu}}$ with:

\begin{equation}
	\bm{\Upsilon}_{\mu\nu} = \frac{1}{2} (\zbar^{*}_{\mu} \zbar_{\nu} + \zbar_{\nu}\zbar^{*}_{\mu})
	\label{eq:ups} = \eta^{\mu \alpha} \aleph_{\alpha\nu} + \frac{1}{2} \eta_{\mu \nu}
\end{equation}

The associated commutation relation is : 

\begin{equation}
	[\bm{\Upsilon}_{\mu\nu}, \bm{\Upsilon}_{\rho\sigma} ] = \eta_{\mu \rho} \bm{\Upsilon}_{\mu\sigma} - \eta_{\nu \sigma}\bm{\Upsilon}_{\rho\nu} 
	\label{eq:comb}
\end{equation}

This commutation relations reflect the existence of an isomorphism between the algebra generated by $\bm{\Upsilon}_{\mu \nu}$ operator and $\mathfrak{u}(1,4)$. Moreover, one can observe similarity  with the commutation relation in (\ref{eq:com}). Given this isomorphism, the following Casimir operators can be deduced: 

\begin{equation}	
	\begin{cases}
		\bm{\mathcal{C}}^{(1)}_{B} = \eta^{\mu \nu} \bm{\Upsilon}_{\mu \nu} \\
		\bm{\mathcal{C}}^{(2)}_{B} =  \eta^{\mu \rho} \eta^{\nu \sigma} \bm{\Upsilon}_{\mu \nu} \bm{\Upsilon}_{\rho \sigma}
	\end{cases}
	\label{eq:bcasimir}
\end{equation}

Like in the \cite{ravo2, ravo5} we introduce the operator:

\begin{equation}
	\bm{\aleph_}{\mu\nu} = \bm{\zbar^{\dagger}_{\mu}} \bm{\zbar_{\nu}}
\end{equation}

The eigenstates of the $\bm{\aleph_}{\mu\mu}$ are the bosonic-like states:

\begin{equation}
	 \lvert n, \langle z \rangle \rangle = \lvert n_0,n_1,n_2,n_3,n_4, \langle z_0\rangle,\langle z_1\rangle ,\langle z_2\rangle ,\langle z_3\rangle ,\langle z_4\rangle  \rangle 
\end{equation}

defined by the relation :

\begin{equation}
	\begin{aligned}
	\lvert n, \langle z \rangle \rangle & =	\prod_{\mu=0}^{D-1}\left(\frac{(\zbar_\mu^\dagger)^{n_\mu}}{\sqrt{n_\mu!}}\right) \lvert \langle z \rangle \rangle \qquad,\hspace{2cm} \text{with}\qquad D = 5 \\
	& =  \Bigg{[} \left(\frac{(\zbar_0^\dagger)^{n_0}}{\sqrt{n_0!}}\right)  \left(\frac{(\zbar_1^\dagger)^{n_1}}{\sqrt{n_1!}}\right) 	\left(\frac{(\zbar_2^\dagger)^{n_2}}{\sqrt{n_2!}}\right)   \left(\frac{(\zbar_3^\dagger)^{n_3}}{\sqrt{n_3!}}\right)  	\left(\frac{(\zbar_4^\dagger)^{n_4}}{\sqrt{n_4!}}\right)  \Bigg{]} \lvert \langle z \rangle \rangle 
	\end{aligned}
	\label{eq:bstates}
\end{equation}

We have between $\bm{\Upsilon_{\mu \nu}}$ and $\bm{\aleph_{\mu \nu}}$ the following relation: 
\begin{equation}
	\begin{cases}
		\bm{\aleph}_{00} &= \bm{\Upsilon}_{00} - \frac12, \\
		\bm{\aleph}_{0j} &= \bm{\Upsilon}_{0j}, \\
		\bm{\aleph}_{j0} &= -\bm{\Upsilon}_{0j}, \\
		\bm{\aleph}_{jj} &= -\bm{\Upsilon}_{jj} + \frac12, \\
		\bm{\aleph}_{jk} &= -\bm{\Upsilon}_{jk} \quad (j\neq k).
	\end{cases}
	\iff
	\begin{cases}
	\bm{\Upsilon}_{00} &=  \bm{\aleph}_{00} + \frac12, \\
	\bm{\Upsilon}_{0j}  &=  \bm{\aleph}_{0j}, \\
	\bm{\Upsilon}_{0j} &= -\bm{\aleph}_{j0}, \\
	\bm{\Upsilon}_{jj}  &= -\bm{\aleph}_{jj}+ \frac12, \\
	\bm{\Upsilon}_{jk}  &= - \bm{\aleph}_{jk}\quad (j\neq k).
   \end{cases}
   \label{eq:alphupsi}	
\end{equation}

From  relations (\ref{eq:bcasimir}), (\ref{eq:bstates}) and (\ref{eq:alphupsi}) we can deduce that the eigenvalues equations of the Casimir operators are: 

\begin{equation}
	\begin{cases}
		\bm{\mathcal{C}}^{(1)}_{B}\lvert n_0,\dots,n_4, \langle z_0\rangle,\dots,\langle z_4\rangle  \rangle  &= \left(N + \frac{5}{2}\right) \lvert n_0,\dots,n_4, \langle z_0\rangle,\dots,\langle z_4\rangle  \rangle , \\
		\bm{\mathcal{C}}^{(2)}_{B} \lvert n_0,\dots,n_4, \langle z_0\rangle,\dots,\langle z_4\rangle  \rangle  &= \left(N^2 + 5N + \frac{5}{4}\right) \lvert n_0,\dots,n_4, \langle z_0\rangle,\dots ,\langle z_4\rangle  \rangle .
	\end{cases}
\end{equation}

The operators $\bm{\mathcal{C}}^{(1)}_{B}$ and $\bm{\mathcal{C}}^{(2)}_{B}$  have the same eigenstates $\lvert n_0,\dots,n_4, \langle z_0\rangle,\dots,\langle z_4\rangle  \rangle $ as the operators  $\bm{\aleph}_{\mu \mu}$ and  $\bm{\aleph}$, and their eigenvalues are respectively $\left(N + \frac{5}{2}\right)$  and  $\left(N^2 + 5N + \frac{5}{4}\right)$.\\

\subsection{Casimir operators for the hybrid representation }

In the framework of the spin representation of the Linear Canonical Transformations (LCTs) associated with the quantum phase space for signature $(1,4)$, a particularly important object emerges: the hybrid operator $\Zbbar$ given in the relation (\ref{eq:hybrid}). This operator bridges the bosonic and fermionic sectors of the theory and plays a central role in the symmetry description of the extended phase space.\\

The square of $\Zbbar$ yields an invariant quadratic operator under LCTs:

\begin{equation}
	(\Zbbar)^2 = \bm{\aleph} + \bm{\Sigma},
\end{equation}

where $\bm{\aleph} = \delta^{\mu\nu} \zbar_\mu^\dagger \zbar_\nu$ is the bosonic invariant and $\bm{\Sigma} = \delta_{\mu\nu} \bm{\zeta}^{\mu^\dagger} \bm{\zeta}^\nu$ is the fermionic invariant. This decomposition shows that $(\Zbbar)^2$ is a mixed bosonic-fermionic invariant operator of the symmetry group $\mathbb{S} \subset \mathrm{Spin}(2,8)$.

The eigenvalues of $(\Zbbar)^2$ on the joint eigenstates $\lvert n, f, \langle z \rangle\rangle$ are:

\begin{equation}
	(\Zbbar)^2 \lvert n, f, \langle z \rangle\rangle = \big( \lvert n\lvert + \lvert f \lvert \big) \lvert n, f, \langle z \rangle\rangle,
\end{equation}

where $\lvert n\lvert = \sum_{\mu=0}^4 n_\mu$ and $\lvert f\lvert = \sum_{\mu=0}^4 f_\mu$ are the total bosonic and fermionic occupation numbers, respectively.\\

From the decomposition $(\Zbbar)^2 = \bm{\aleph} + \bm{\Sigma}$, and using the known Casimir operators of the bosonic and fermionic sectors, we can define the hybrid Casimir operators.

The  linear hybrid Casimir is simply the sum of the linear bosonic and fermionic Casimirs:

\begin{equation}
	\bm{\mathcal{C}}^{(1)}_{\mathrm{hyb}} = \bm{\mathcal{C}}^{(1)}_B + \bm{\mathcal{C}}^{(1)}_F = \left( \aleph + \frac{5}{2} \right) + \left( \boldsymbol{\Sigma} - \frac{5}{2} \right) = \aleph + \boldsymbol{\Sigma}
\end{equation}

Thus:

\begin{equation}
	\bm{\mathcal{C}}^{(1)}_{\mathrm{hyb}} = (\Zbbar)^2
\end{equation}

Its eigenvalue on the joint eigenstates $|n, f, \langle z \rangle\rangle$ is:

\begin{equation}
	\mathcal{C}^{(1)}_{\mathrm{hyb}} |n, f, \langle z \rangle\rangle = \big( |n| + |f| \big) |n, f, \langle z \rangle\rangle.
\end{equation}

The \textbf{quadratic hybrid Casimir} can be constructed from the quadratic Casimirs of the bosonic and fermionic sectors. Using earlier results:

\begin{equation}
	\bm{\mathcal{C}}^{(2)}_B = \bm{\aleph}(\bm{\aleph} + 5) + \frac{5}{4}, \quad
	\bm{\mathcal{C}}^{(2)}_F = \frac{5}{4}.
\end{equation}

A natural symmetric combination gives:

\begin{equation}
	\mathcal{C}^{(2)}_{\mathrm{hyb}} = \bm{\mathcal{C}}^{(2)}_B + \bm{\mathcal{C}}^{(2)}_F = \aleph(\aleph + 5) + \frac{5}{2}.
\end{equation}

Its eigenvalue on $|n, f, \langle z \rangle\rangle$ is:

\begin{equation}
	\mathcal{C}^{(2)}_{\mathrm{hyb}} |n, f, \langle z \rangle\rangle = ( \lvert n \lvert (\lvert n \lvert + 5) + \frac{5}{2}) |n, f, \langle z \rangle\rangle.
\end{equation}

Remarkably , the fermionic part of $	\mathcal{C}^{(1)}_{\mathrm{hyb}} = \Zbbar$ is directly related to the charge operators of the Standard Model. Indeed, the operators $\boldsymbol{\Sigma}^{\mu\mu} = \boldsymbol{\zeta}^{\mu^\dagger} \boldsymbol{\zeta}^\mu$ allow the identification \cite{ravo2, ravo5}:

\begin{equation}
	I_3 = \frac12(\boldsymbol{\Sigma}^{00} + \boldsymbol{\Sigma}^{44}) - \frac12, \quad
	Y_W = \boldsymbol{\Sigma}^{00} - \frac23(\boldsymbol{\Sigma}^{11}+\boldsymbol{\Sigma}^{22}+\boldsymbol{\Sigma}^{33}) - \boldsymbol{\Sigma}^{44} + 1, \quad
	Q = I_3 + \frac{Y_W}{2}.
\end{equation}

Thus, the hybrid operators not only unifies bosonic and fermionic invariants but also encodes standard model charges structures within its algebraic formulation. They may offer a pathway to understand mass hierarchies, generation replication and the origin of neutrinos masses within a unified phase space symmetry framework.

\section{Discussion and Conclusion}

In this work, we have identified linear and quadratic Casimir operators for the symmetry group of the relativistic quantum phase space corresponding to the signature $(1,4)$ by using a unified framework based on the spin representation of this group which is formed by Linear Canonical Transformations (LCTs).\\

A significant conceptual outcome of our approach is the natural unification of internal and spacetime symmetries within the framework of relativistic quantum phase space \cite{ravo1}. In conventional quantum field theory, internal symmetries (such as $SU(2)_L \times U(1)_Y$) and spacetime symmetries (such as the Poincaré group) are treated separately, a distinction enforced by the Coleman-Mandula theorem under certain hypotheses \cite{coleman1967all, Oskar}. However, the quantum phase space formalism, through its associated LCT group $\mathbb{T}$, extends the notion of spacetime symmetry to include transformations that mix momenta and coordinates in a way that is compatible with the uncertainty principle. The spin representation of $\mathbb{T}$ further incorporates fermionic degrees of freedom via Clifford algebras, thereby embedding what are traditionally viewed as internal quantum numbers into a geometric quantum phase space structure. This embedding suggests a pathway to transcend the limitations imposed by the Coleman-Mandula theorem, which assumes a purely spacetime Poincaré symmetry. In our framework, the approach aligns with more general no go theorem evasion strategies, such as those in supersymmetry or conformal field theory, but here arises from first principles of relativistic quantum phase space covariance.\\

From a mathematical perspective, our construction connects to the theory of non-compact group representations and has similarities with the work of Harish-Chandra on semisimple Lie groups \cite{harish} and Mackey’s theory of induced representations \cite{mackey}. In fact, the group $\mathbb{T} \cong Sp(2,8)$ is a non-compact group, and its representation theory is less standard than that of compact groups. While Harish-Chandra’s approach focuses on the decomposition of $L^2(G)$ via the Plancherel theorem, and Mackey’s method builds representations from subgroups, our approach is more algebraic and constructive: we directly build Fock-like representations via bosonic and fermionic ladder operators $\zbar_\mu, \boldsymbol{\zeta}^\mu$ derived from the quantum phase space structure. This allows explicit state labeling by occupation numbers $(n_\mu, f^\mu)$ and direct computation of Casimir eigenvalues. Compared to the full-blown Harish-Chandra theory, our method is computationally more accessible and physically transparent, especially for the purpose of particle classification and connection with the standard model.\\

However, open question remains concerning, for instance, the physical interpretation of the bosonic occupation numbers $n_\mu$ appearing in the joint eigenstates $|n,f,\langle z\rangle\rangle$. While $f^{\mu}$ determines Standard Model charges, $n_{\mu}$ are so far unassigned to known quantum numbers. We speculate that $|n|$ or perhaps finer structures within the $n_\mu$ could label fermion generations or relate to mass hierarchy patterns. This would align with the fact that sterile neutrinos appear in three flavors, each possibly associated with distinct bosonic backgrounds. Moreover, the hybrid Casimir eigenvalues $\mathcal{C}^{(1)}_{\mathrm{hyb}} = |n| + |f|$ and $\mathcal{C}^{(2)}_{\mathrm{hyb}} = |n| (|n| + 5) + 5/2$ suggest a combined excitation number that may be linked to mass or other fundamental particle properties  in a deeper dynamical framework. \\

To conclude, we can say that the geometric approach via the relativistic quantum phase space offers a fresh perspective on the origin of gauge quantum numbers, particle properties and generation replication. The quantum phase space symmetry framework not only explains the possible existence of sterile neutrinos but also points toward a deeper geometric unification of symmetry principles in fundamental physics. Future research directions may include: coupling the quantum numbers $n_\mu$ and $f^{\mu}$ to known physical observables (masses, mixing angles),extending the formalism to include interactions and dynamics, possibly through a phase space gauge principle and exploring cosmological implications, given that the $(1,4)$ signature is naturally linked to de Sitter spacetime and a positive cosmological constant.

\end{document}